\documentclass[a4paper,11pt]{article} 
\pdfoutput=1
\usepackage{graphicx,rotating,hyperref,slashed,amsmath,xcolor,amssymb,amsfonts,colortbl,cite, subfigure,float} 
\makeatletter 
\usepackage{graphics} 
\usepackage{graphicx} 
\graphicspath{{plots/}}
 
\hypersetup{colorlinks,bookmarksopen,bookmarksnumbered,
linkcolor=blus,pdfstartview=FitH,urlcolor=rossos,citecolor=verde}
\numberwithin{equation}{section}
\usepackage{amsmath, tcolorbox}
\hypersetup{%
    ,urlcolor=rossos
    ,citecolor=rossos
    ,linkcolor=rossos
    }

\AtBeginDocument{
  \hypersetup{
    urlcolor=verdes,
    citecolor=verdes,
    linkcolor=verdes,
  }%
}

\allowdisplaybreaks

\def\lsim{\mathrel{\rlap{\lower3pt\hbox{\hskip0pt$\sim$}}
   \raise1pt\hbox{$<$}}}         
\def\gsim{\mathrel{\rlap{\lower4pt\hbox{\hskip1pt$\sim$}}
   \raise1pt\hbox{$>$}}}         

 \newcommand{\sfootnote}[1]{} 
\definecolor{bluc}{cmyk}{1,1,0,0.1}
\definecolor{rossoCP3}{cmyk}{0,.88,.77,.40}
\definecolor{rosso}{cmyk}{0,1,1,0.4}
\definecolor{rossos}{cmyk}{0,1,1,0.55}
\definecolor{rossoc}{cmyk}{0,1,1,0.2}
\definecolor{verdes}{cmyk}{0.92,0,0.59,0.4}

\hypersetup{colorlinks, bookmarksopen, bookmarksnumbered,
citecolor=verdes, linkcolor=bluc, pdfstartview=FitH, urlcolor=rossos}

\newcommand{\mio}[1]{}

\definecolor{Gray}{gray}{0.95}

\usepackage{multicol}
\usepackage{color}
\definecolor{rosso}{cmyk}{0,1,1,0.4}
\definecolor{rossos}{cmyk}{0,1,1,0.55}
\definecolor{rossoc}{cmyk}{0,1,1,0.2}
\definecolor{blu}{cmyk}{1,1,0,0.3}
\definecolor{blus}{cmyk}{1,1,0,0.6}
\definecolor{bluc}{cmyk}{1,1,0,0.1}
\definecolor{verde}{cmyk}{0.92,0,0.59,0.25}
\definecolor{verdec}{cmyk}{0.92,0,0.59,0.15}
\definecolor{verdes}{cmyk}{0.92,0,0.59,0.4}

\def\circa#1{\,\raise.3ex\hbox{$#1$\kern-.75em\lower1ex\hbox{$\sim$}}\,}

\newcommand{\beq}{\begin{equation}}
\newcommand{\eeq}{\end{equation}}

\newcommand{\bea}{\begin{eqnarray}}
\newcommand{\eea}{\end{eqnarray}}
\newcommand{\be}{\begin{equation}}
\newcommand{\ee}{\end{equation}}

\newfam\rsfsfam
\def\mathscr#1{{\fam\rsfsfam\relax#1}}

\def\circa#1{\,\raise.3ex\hbox{$#1$\kern-.75em\lower1ex\hbox{$\sim$}}\,}
\makeatletter

\def\hhref#1{\href{http://arxiv.org/abs/#1}{arXiv:#1}} 

\newcommand{\doi}[1]{\href{http://dx.doi.org/#1}{[doi]}}

\def\hhref#1{\href{http://arxiv.org/abs/#1}{arXiv:#1}} 
 
\def\art{\@ifnextchar[{\eart}{\oart}}
\def\eart[#1]#2#3#4#5#6{{\rm #2}, {\em #3 \bf #4} {\rm (#6) #5} ({\em #1})}

\def\article{\@ifnextchar[{\earticle}{\oarticle}}
\def\oarticle#1#2#3#4#5#6{{\rm #1}, {\em ``#6''}, {\rm #2 #3 (#5) #4}}
\def\earticle[#1]#2#3#4#5#6#7{{\rm #2}, {\em ``#7''}, {\rm #3 #4 (#6) #5}  [\hhref{#1}]}
\def\hepart[#1]#2{{\rm #2, \em#1}}
\def\heparticle[#1]#2#3{#2, {\em ``#3''} [\hhref{#1}]}

\newcounter{alphaequation}[equation]
\def\thealphaequation{\theequation\hbox to
0.6em{\hfil\alph{alphaequation}\hfil}}
\def\eqnsystem#1{
\def\@eqnnum{{\rm (\thealphaequation)}}
\def\@@eqncr{\let\@tempa\relax \ifcase\@eqcnt \def\@tempa{& & &} \or
  \def\@tempa{& &}\or \def\@tempa{&}\fi\@tempa
  \if@eqnsw\@eqnnum\refstepcounter{alphaequation}\fi
\global\@eqnswtrue\global\@eqcnt=0\cr}
\refstepcounter{equation} \let\@currentlabel\theequation \def\@tempb{#1}
\ifx\@tempb\empty\else\label{#1}\fi
\refstepcounter{alphaequation}
\let\@currentlabel\thealphaequation
\global\@eqnswtrue\global\@eqcnt=0 \tabskip\@centering\let\\=\@eqncr
$$\halign to \displaywidth\bgroup \@eqnsel\hskip\@centering
$\displaystyle\tabskip\z@{##}$&\global\@eqcnt\@ne
\hskip2\arraycolsep\hfil${##}$\hfil& \global\@eqcnt\tw@\hskip2\arraycolsep
$\displaystyle\tabskip\z@{##}$\hfil
\tabskip\@centering&\llap{##}\tabskip\z@\cr}
\def\endeqnsystem{\@@eqncr\egroup$$\global\@ignoretrue} \makeatother

\definecolor{fiorentina}{rgb}{.5,0,.5}

\begin{document}

\vspace{1truecm}
\begin{center}
\boldmath

 \huge{ \bf
{\boldmath \(f(R)\) Gravity: Gravitational Waves Tests}}

\unboldmath
\end{center}
\unboldmath

\vspace{-0.2cm}

\begin{center}
{\fontsize{12}{30}\selectfont  
Rafid H. Dejrah \footnote{\texttt{rafid.dejrah@gmail.com}}}

\end{center}

\begin{center}

\vskip 8pt
\textsl{Department of Physics, Faculty of Sciences, Ankara University,  06100, Ankara, Türkiye}\\

\vskip 7pt

\end{center}
\smallskip
\begin{center}
\begin{abstract}
\noindent 
This review explores modified theories of gravity, particularly \(f(R)\) gravity, as extensions to General Relativity (GR) that offer alternatives to dark energy for explaining cosmic acceleration. These models generalize the Einstein-Hilbert action to include functions of the Ricci scalar, providing new insights into cosmology and astrophysics. The detection of gravitational waves (GWs) has enabled rigorous tests of \(f(R)\) gravity, as deviations in GW propagation, speed, and polarization can signal modifications to GR. Constraints on \(f(R)\) models arise from LIGO-Virgo observations of binary mergers, the stochastic gravitational wave background (SGWB), and complementary tests in cosmology and weak-field regimes. Future GW detectors, such as LISA and the Einstein Telescope, will enhance sensitivity to smaller deviations from GR, necessitating advancements in theoretical modeling. Among competing theories---including scalar-tensor, massive gravity, and Horndeski models---\(f(R)\) gravity remains a pivotal framework for understanding fundamental gravitational physics and cosmology. This review highlights key developments, challenges, and future directions in the field.
\end{abstract}
\end{center}

\newpage
\tableofcontents
\section{Introduction}
\label{sec:introduction}
General Relativity (GR), formulated by Albert Einstein has been the cornerstone of modern gravitational theory. The framework of GR describes gravity as the curvature of spacetime caused by mass and energy, encapsulated in the well-known Einstein field equations (EFE):
\begin{equation}
    R_{\mu \nu} - \frac{1}{2} g_{\mu \nu} R + g_{\mu \nu} \Lambda = \frac{8 \pi G}{c^4} T_{\mu \nu},
\end{equation}
where \( R_{\mu \nu} \) is the Ricci curvature tensor, \( g_{\mu \nu} \) is the metric tensor, \( \Lambda \) is the cosmological constant \cite{Weinberg_1989, weinberg2000cosmologicalconstantproblemstalk}, and \( T_{\mu \nu} \) is the stress-energy tensor. While GR has successfully passed a multitude of experimental tests \footnote{See Table 1 in \cite{Shankaranarayanan_2022} for a summary of experimental tests of GR.} in the weak-field regime \cite{Will_2014}, it faces significant challenges in explaining certain cosmological phenomena, such as the accelerated expansion of the Universe \cite{Clifton_2012}. This has motivated the exploration of modified gravity theories, which extend the classical framework to address these issues without the need for dark energy.

Despite its successes, GR is not without its limitations. One of the most pressing challenges is the observed accelerated expansion of the Universe, first discovered through observations of Supernovae Type Ia  (SNe Ia) \cite{Riess_1998, Perlmutter_1999, Chickles:2024qil}. This phenomenon is typically attributed to dark energy, a mysterious component of the Universe that exerts negative pressure and drives the acceleration. However, the nature of dark energy remains one of the greatest unsolved problems in modern physics. The cosmological constant \(\Lambda\), introduced by Einstein to achieve a static universe, is the simplest explanation for dark energy within the framework of GR. However, the observed value of \(\Lambda\) is many orders of magnitude smaller than predictions from quantum field theory, leading to the so-called ``cosmological constant problem'' \cite{Weinberg_1989, weinberg2000cosmologicalconstantproblemstalk}.

Another challenge to GR arises from the need to explain the observed dynamics of galaxies and galaxy clusters without invoking dark matter. While dark matter provides a successful explanation for the rotation curves of galaxies and the gravitational lensing of light, its particle nature remains elusive despite decades of experimental searches \cite{Bertone_2005}. These challenges have led to the exploration of modified gravity theories, which seek to explain these phenomena by altering the laws of gravity on large scales.

\subsection{Modified Gravity Theories}
\label{subsec:modified_gravity}

Modified gravity theories aim to address the shortcomings of GR by introducing new degrees of freedom or modifying the gravitational action. Among the various alternatives to GR, \( f(R) \) gravity theories \cite{Sotiriou_2010} have emerged as a compelling class of models \cite{Geng:2012zc}. These theories modify the Einstein-Hilbert action by replacing the Ricci scalar \( R \) with an arbitrary function \( f(R) \), leading to modified field equations that govern the dynamics of spacetime. This modification can account for cosmic acceleration and other phenomena \cite{Appleby_2007, Nojiri_2011} typically attributed to dark energy \cite{Tsujikawa_2008}. Moreover, \( f(R) \) gravity is interesting because it introduces additional scalar degrees of freedom, which can manifest as new observable effects, such as modifications to the propagation of gravitational waves (GWs).

The action for \( f(R) \) gravity is given by:
\begin{equation} \label{f(R)_action}
    S = \int d^4x \sqrt{-g} \left( \frac{1}{2\kappa} f(R) + \mathcal{L}_m \right),
\end{equation}
where \( \kappa = 8 \pi G \), and \( \mathcal{L}_m \) is the matter Lagrangian. The function \( f(R) \) defines the dynamics of spacetime, and the field equations derived from this action can lead to different solutions in the weak-field regime. This is particularly important for the study of gravitational wave propagation, as any modifications to the gravitational action will alter the speed, polarization, and other characteristics of GWs.

\subsection{Gravitational Waves as a Probe of Modified Gravity}
\label{subsec:gw_probe}

The direct detection of gravitational waves by the LIGO and Virgo collaborations \cite{Abbott_2016} has opened a new window into the Universe, providing a powerful tool for testing GR and its alternatives. GWs are ripples in spacetime caused by the acceleration of massive objects, such as merging black holes or neutron stars. In GR, GWs propagate at the speed of light and have two tensor polarization modes \cite{Isi_2017}: \( h_+ \) and \( h_\times \). However, in modified gravity theories like \( f(R) \) gravity, additional scalar polarization modes can arise, and the propagation speed of GWs may differ from the speed of light.

The observation of GW170817 \cite{Abbott:2017}, a binary neutron star merger accompanied by a gamma-ray burst, provided a stringent test of the propagation speed of GWs. The time delay between the GW and electromagnetic signals constrained the deviation of the GW speed from the speed of light to:
\begin{equation}
\frac{|v_{\text{GW}} - c|}{c} \lesssim 10^{-15}.
\end{equation}
This result ruled out many modified gravity models that predict significant deviations in the GW propagation speed.

In Section \ref{f(R)}, we provide a detailed discussion of modified \( f(R) \) gravity theories. In Section \ref{sec:constraints}, we examine potential constraints on \( f(R) \) gravity models based on gravitational wave measurements. Next, in Section \ref{sec:future}, we explore future constraints on both the theoretical framework of \( f(R) \) gravity and its implications for cosmological observations. Other modified gravity theories are discussed in Section \ref{sec:other_theories}, where we present a detailed analysis of each. Finally, in Section \ref{sec:conclusions}, we summarize our discussions and offer remarks on the future prospects of testing \( f(R) \) gravity models through gravitational waves.


\section{$f(R)$ Gravity}
\label{f(R)}
Modified gravity theories are an essential part of the modern effort to extend GR, \cite{Wands_1994, Capozziello_2011}, to account for phenomena such as cosmic acceleration and dark energy \cite{Amendola:2012}. Among the various alternatives to GR, \( f(R) \) gravity theories, in which the Einstein-Hilbert action is generalized by replacing the Ricci scalar \( R \) with an arbitrary function \( f(R) \), have gained significant attention. In this section, we will discuss the theoretical foundations of \( f(R) \) gravity, the state of current research, and the contributions made by various researchers in this field.

The fundamental action for \( f(R) \) gravity is given by Eq.\eqref{f(R)_action}. The function \( f(R) \) characterizes the modification of gravity, with the simplest choice being \( f(R) = R \), which recovers GR. However, by choosing other forms of \( f(R) \), one can explore the impact of modifications to gravity, such as the introduction of new scalar degrees of freedom. These modifications can have significant consequences for cosmology, gravitational wave propagation, and the formation of structure in the Universe.

A widely studied example of \( f(R) \) gravity is the Starobinsky model \cite{Starobinsky:1980te}, which introduces a quadratic correction to the Ricci scalar:
\begin{equation}\label{Starobinsky}
    f(R) = R + \alpha R^2,
\end{equation}
where \( \alpha \) is a constant that determines the strength of the modification. Other forms of \( f(R) \) gravity include exponential functions, such as:
\begin{equation}
    f(R) = R + \beta e^{\gamma R},
\end{equation}
which are often used to model inflationary scenarios. These models are capable of generating late-time cosmic acceleration by modifying the gravitational dynamics without the need for dark energy. Other forms of \( f(R) \) include exponential models, which are often used to explain inflationary dynamics \cite{Cognola_2008}, and models involving logarithmic corrections. These modifications are expected to lead to deviations in the behavior of gravitational waves, especially in the high-energy regime where the effects of the scalar degrees of freedom become significant.

\subsection*{Key Contributions to the Field}

The study of \( f(R) \) gravity has evolved significantly over the past few decades, with numerous contributions from theoretical physicists working on both the mathematical formulation and observational constraints of these models. Below, we provide a summary of key areas of research and prominent contributors to each:

\subsubsection*{Cosmological Applications of $f(R)$ Gravity}

The modification of gravity at cosmological scales is one of the main motivations for the study of \( f(R) \) gravity. Many researchers have worked on using \( f(R) \) gravity to explain the observed accelerated expansion of the Universe. The most notable of these is the work by Starobinsky, who introduced the quadratic model, Eq. \eqref{Starobinsky}, to explain cosmic acceleration without invoking dark energy. This model is widely studied in the context of inflationary cosmology \cite{Guth:1980zm}, as it naturally leads to a de Sitter-like solution at high curvature, which can account for the observed expansion at early times \cite{NOJIRI20171}.

In addition to Starobinsky's work, several researchers have explored different forms of \( f(R) \) gravity to address the late-time acceleration of the Universe. For example, the model by \cite{Carroll_2004} introduced an \( f(R) \) model that leads to a self-accelerating cosmology, without the need for dark energy. This work was influential in showing that modifications of gravity could provide an alternative explanation for accelerated cosmic expansion.

Another key direction of research in this area has been the study of the effective equation of state in \( f(R) \) gravity. This work by \cite{PhysRevD.68.123512} has been instrumental in deriving the cosmological field equations and understanding the behavior of the equation of state in various \( f(R) \) models. These studies have shown that different forms of \( f(R) \) gravity can lead to different accelerations at late times, depending on the specific form of the function \( f(R) \).

\subsubsection*{Gravitational Wave Propagation in $f(R)$ Gravity}

One of the key predictions of \( f(R) \) gravity is that gravitational waves may propagate differently than in GR due to the additional scalar degree of freedom. The modification of the gravitational action in \( f(R) \) gravity can alter the speed of gravitational waves and introduce new polarization states \cite{chowdhury2021gravitationalwavesmodifiedgravity}, which could be detected by current and future gravitational wave observatories.

In the context of GW propagation, a significant contribution was made by \cite{Ananda:2007xh, Katsuragawa_2019, chowdhury2021gravitationalwavesmodifiedgravity, zhou2024scalarinducedgravitationalwaves}, who analyzed the propagation of gravitational waves in the \( f(R) \) framework and derived the modified field equations for the perturbations of the metric. Their work showed that in the presence of \( f(R) \) modifications, gravitational waves could propagate at speeds different from the speed of light, potentially leading to observable effects in GW observations from binary mergers and other astrophysical sources.

Further research by \cite{Geng:2012zc} extended this analysis, focusing on the perturbative approach to gravitational waves in \( f(R) \) gravity, and showed how the scalar modes could affect the polarization of gravitational waves. In \( f(R) \) gravity, the additional scalar degree of freedom, often referred to as the ``scalaron'', modifies the propagation of gravitational waves by introducing a new polarization mode. This scalar mode arises from the fact that \( f(R) \) gravity can be recast as a scalar-tensor theory, where the Ricci scalar \( R \) is coupled to a scalar field. The presence of this scalar mode leads to deviations from the standard tensor polarizations (\( h_+ \) and \( h_\times \)) predicted by General Relativity (GR), providing a unique signature that can be tested with gravitational wave detectors.

The perturbative approach to gravitational waves in \( f(R) \) gravity involves analyzing the linearized field equations around a background metric. In this framework, the scalar mode manifests as a longitudinal component of the gravitational wave, which is absent in GR. This scalar mode can affect the amplitude, phase, and propagation speed of gravitational waves, leading to observable differences in the detected signals. For example, \cite{Gong:2018ybk} demonstrated that the scalar mode could induce a frequency-dependent modification to the gravitational wave dispersion relation, which could be detected as a phase shift in the waveform. Additionally, \cite{Saltas:2014dha} showed that the scalar mode could alter the energy spectrum of the stochastic gravitational wave background (SGWB), providing another avenue for testing \( f(R) \) gravity.

These modifications are crucial for testing \( f(R) \) gravity with detectors like LIGO and Virgo, as they provide a means to distinguish between different gravity models based on the behavior of the observed GW signals. For instance, the absence of scalar polarization modes in the detected signals from binary black hole mergers has already placed stringent constraints on \( f(R) \) gravity models \cite{Abbott:2017}. Future observations with next-generation gravitational wave detectors, such as the Einstein Telescope and Cosmic Explorer, will significantly enhance our ability to test modifications to general relativity. These advanced instruments will offer unprecedented sensitivity to key properties of gravitational waves, including their propagation speed and polarization content. In particular, multi-messenger observations---such as the coincident detection of gravitational waves and their electromagnetic counterparts---enable precise constraints on the propagation speed of gravitational waves, a test originally proposed by~\cite{Nishizawa:2014zna,Nishizawa:2016kba}. In the context of general modified gravity theories, such as certain formulations of Horndeski gravity, the tensor modes of gravitational waves may propagate at speeds different from that of light. Current observations have already placed stringent limits on such deviations, as demonstrated in Ref.~\cite{Nishizawa:2018}, which specifically constrains the propagation speed of tensor modes in Horndeski theory---a framework that includes \( f(R) \) gravity as a special case.

In \( f(R) \) gravity, both the tensor and scalar modes propagate at the speed of light, consistent with general relativity and with more general modified gravity theories that feature non-minimal coupling between curvature and matter~\cite{Varela:2024egg}. Consequently, deviations in propagation speed do not serve as a distinguishing feature between \( f(R) \) gravity and general relativity. Instead, the hallmark of \( f(R) \) gravity lies in the presence of an additional scalar degree of freedom, which manifests as an extra polarization mode in gravitational waves, beyond the standard plus and cross polarizations of general relativity. This scalar mode, often referred to as a transverse breathing mode, can be probed through detailed polarization analysis of gravitational wave signals. Recent studies utilizing data from the LIGO and Virgo observatories have begun to constrain the amplitude of such scalar modes, providing observational insights into their potential presence~\cite{Takeda:2021hgo,Takeda:2023wqn}.

However, recent advancements have refined our understanding of the linear degrees of freedom in \( f(R) \) gravity, particularly concerning the propagation of gravitational waves in maximally-symmetric space-times, such as flat Minkowski space. Initially, it was assumed that all \( f(R) \) models propagate three polarization modes—two tensor modes associated with the graviton and one scalar mode linked to the scalaron. Contrarily, a significant class of \( f(R) \) models, newly classified as ``degenerate models'', has been shown to possess an empty linear spectrum in such backgrounds, meaning they do not propagate any gravitational waves at all~\cite{Casado-Turrion:2023rni,Casado-Turrion:2024esi}. This class includes notable examples such as \( f(R) = R^2 \) and various power-law forms frequently utilized in cosmological applications. These findings~\cite{Casado-Turrion:2023rni,Casado-Turrion:2024esi} overturn the earlier belief of universal propagation of three polarizations across all \( f(R) \) theories, highlighting that the presence of gravitational wave modes is contingent upon the specific functional form of \( f(R) \). Moreover, these degenerate models exhibit stability issues and other unphysical properties, casting doubt on their physical viability as consistent gravitational theories~\cite{Casado-Turrion:2023rni,Casado-Turrion:2024esi}. This revelation underscores the necessity of carefully delineating between degenerate and non-degenerate \( f(R) \) models when interpreting gravitational wave observations, as only the latter support the additional scalar polarization mode previously discussed.

In summary, the perturbative approach to gravitational waves in \( f(R) \) gravity highlights the importance of scalar modes as a distinguishing feature of modified gravity theories. By analyzing the polarization, dispersion, and energy spectrum of gravitational waves, we can test the predictions of \( f(R) \) gravity and constrain its parameters. These studies not only advance our understanding of gravity but also provide a powerful tool for probing the fundamental nature of the Universe.

Another important area of research in \( f(R) \) gravity is the study of weak-field limits and solar system tests. Since GR has been experimentally verified in the weak-field regime through experiments such as the deflection of light and the precession of Mercury's orbit \cite{Park_2017}, any modification to gravity must satisfy these stringent constraints.

\subsubsection*{Solar System Tests and Weak-Field Constraints}
Recent advancements in cosmology have led to significant developments in alternative gravity models, challenging the standard assumption that GR holds on all scales. While GR has been rigorously tested within the Solar System, the discovery of the Universe’s accelerated expansion has prompted new questions about its validity at cosmological distances. This acceleration, often attributed to dark energy within the $\Lambda$CDM framework \cite{2020}, could instead suggest that GR itself requires modification on large scales. To explore this possibility, numerous modified gravity theories, such as $f(R)$ gravity, Horndeski theory, and massive gravity, have been proposed. These models introduce novel mechanisms to reconcile deviations from GR while passing Solar System tests through screening effects. In this context, \cite{Koyama_2016} provides a comprehensive review of recent progress in constructing modified gravity models and developing cosmological tests to distinguish them from standard GR-based cosmology.

In recent years, various modifications to gravity have been proposed as alternatives to the dark matter hypothesis \cite{Milgrom:1983ca, Sanders_2002,Bekenstein:2004ne, Clifton_2012, Verlinde_2017}. These modified gravity theories introduce additional apparent force terms in the dynamical equations, which aim to reproduce the observed gravitational effects without invoking unseen mass components. Typically, such extra force terms scale with the radial coordinate, becoming significant on galactic or cosmological scales to account for the missing mass problem. However, their effects may still manifest within smaller-scale structures, such as the solar system. A detailed investigation of these effects has been carried out in \cite{Chan_2022}, where they analytically derive general expressions for the contribution of modified gravity theories to planetary precession in the solar system. The study examines three popular models—Modified Newtonian Dynamics (MOND), Emergent Gravity (EG), and Modified Gravity (MG)—and places constraints on MOND interpolating functions using solar system data.

Works like \cite{Chiba_2003} have developed methods to study the weak-field limits of \( f(R) \) gravity, showing that for most forms of \( f(R) \), the modifications to GR are negligible in the solar system. This is due to the so-called Chameleon Mechanism \footnote{Check Sec. 3.5 in \cite{Erickcek:2009sda} for detailed discussion about the mechanism.} \cite{Brax_2008, Brax_2017, OShea:2024jjw}, which ensures that the scalar degree of freedom in \( f(R) \) gravity is effectively screened in high-density environments like the Earth and Sun. The development of these screening mechanisms has been crucial for ensuring that \( f(R) \) gravity is consistent with solar system observations.

\subsubsection*{The Chameleon Mechanism}

Modified gravity theories, such as $f(R)$ gravity, often introduce additional scalar degrees of freedom. A major challenge is ensuring that these scalar fields remain undetectable in high-density environments, such as the Solar System, while still influencing cosmological dynamics. The Chameleon Mechanism provides a solution by allowing the scalar field to have a mass that depends on the local matter density. This ensures that modifications to gravity are screened in dense regions but manifest in low-density environments such as cosmic voids.

In $f(R)$ gravity, the action is given by Eq. \eqref{f(R)_action}. This can be rewritten in a scalar-tensor form by introducing an auxiliary field \( \phi \), defined as:
\begin{equation}
\phi = f_R \equiv \frac{df}{dR}.
\end{equation}
\begin{figure}[t!]
    \centering
    \includegraphics[width=0.8\textwidth]{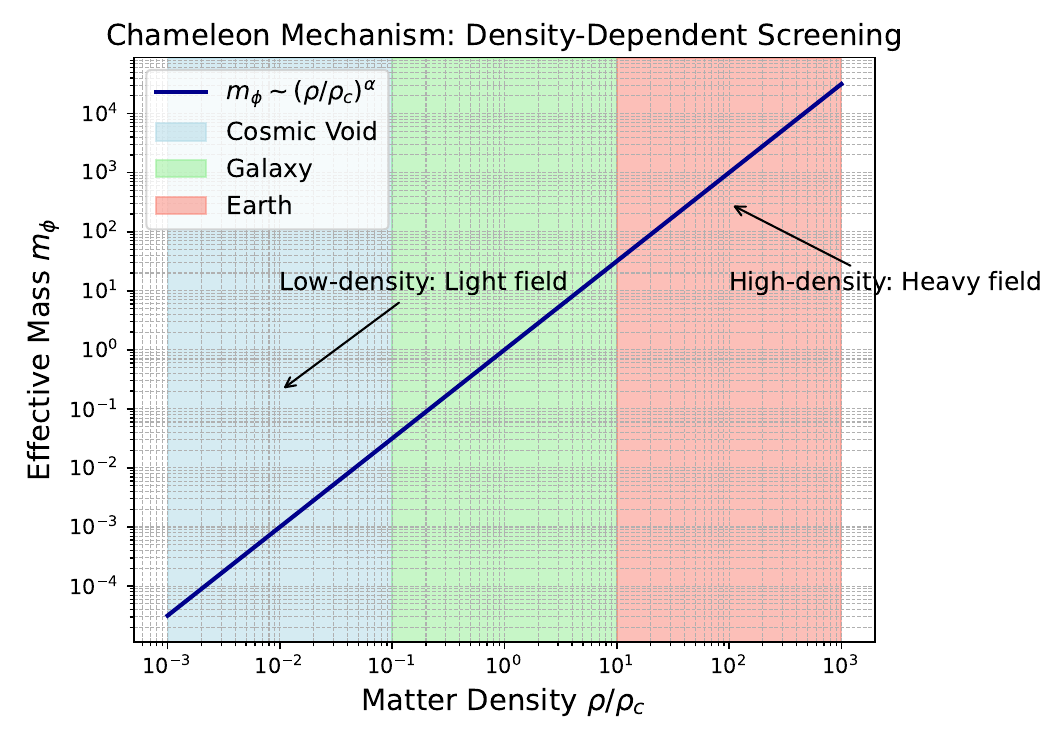}
    \caption{The effective mass of the chameleon scalar field (\(m_\phi\)) as a function of the local matter density (\(\rho\)). The shaded regions indicate different astrophysical environments where the mechanism operates.}
    \label{fig:chameloen}
\end{figure}
The equation of motion for \( \phi \) is:
\begin{equation}
\square \phi = \frac{dV_{\text{eff}}}{d\phi},
\end{equation}
where the effective potential is \cite{Khoury_2004}:
\begin{equation}
V_{\text{eff}}(\phi) = V(\phi) + \rho_m e^{\beta \phi},
\end{equation}
where \( V(\phi) \) is the potential governing the scalar field dynamics, and \( \rho_m \) is the local matter density.

The key feature of the Chameleon Mechanism is the density-dependent mass:
\begin{equation}
m_\phi^2 = \frac{d^2 V_{\text{eff}}}{d\phi^2}.
\end{equation}
\begin{itemize}
    \item \textbf{High-density regions} (e.g., the Earth, Solar System): \( m_\phi \) is large, meaning the field is short-ranged, suppressing deviations from GR.
    \item \textbf{Low-density regions} (e.g., cosmic voids): \( m_\phi \) is small, allowing the field to mediate long-range fifth forces.
\end{itemize}

The Chameleon Mechanism is a screening mechanism in modified gravity theories that allows scalar fields to evade local experimental constraints while remaining cosmologically relevant. The key feature of this mechanism is that the effective mass of the scalar field depends on the local matter density: in high-density regions, such as within the Solar System, the field acquires a large mass, making its interaction range extremely short and suppressing deviations from GR. Conversely, in low-density cosmic environments, such as voids, the field remains light, allowing modifications to gravity to manifest over astrophysical scales. Figure~\ref{fig:chameloen} illustrates this behavior by showing how the Chameleon field’s mass increases with ambient density. The transition between different astrophysical environments highlights the environmental dependence of the screening, with strong suppression in high-density regions (e.g., Earth) and significant effects in low-density regions (e.g., intergalactic space). This behavior is crucial for understanding the viability of Chameleon-based modified gravity models and their observational signatures in gravitational wave experiments and large-scale structure surveys. The Chameleon Mechanism significantly impacts the formation of structures in the Universe:
\begin{itemize}
    \item \textbf{Cosmic Large-Scale Structure:} The mechanism modifies gravitational clustering in cosmic voids, leading to scale-dependent growth.
    \item \textbf{Weak Lensing:} Deviations from GR appear as modifications to the lensing potential.
    \item \textbf{Cosmic Microwave Background (CMB)} Chameleon fields can alter the Integrated Sachs-Wolfe (ISW) effect.
\end{itemize}

Observations and experiments place stringent constraints on Chameleon fields. The Cassini spacecraft provided a Shapiro time delay \cite{PhysRevLett.13.789} constraint \cite{article}:
\begin{equation}
\gamma - 1 = (2.1 \pm 2.3) \times 10^{-5}.
\end{equation}
This limits the deviation from GR induced by the scalar field. Additionally, experiments such as the Eöt-Wash experiment and atom interferometry tests \cite{Schlamminger_2008} probe deviations in Newton’s law. The force mediated by \( \phi \) obeys:
\begin{equation}
F_\phi \propto \frac{e^{-m_\phi r}}{r^2}.
\end{equation}
Current experiments limit \( m_\phi \) in the range [\(10^{-3}\) - \(10^{-2}\)] eV \cite{Brax_2019}.

The Chameleon Mechanism is a compelling way to reconcile modified gravity with local tests of GR. It ensures that scalar fields are screened in high-density environments while allowing deviations in cosmic voids. Future experiments, particularly in cosmology and laboratory physics, will provide deeper insights into the nature of scalar-tensor gravity theories.


\section{Constraints on $f(R)$ Gravity from Gravitational Wave Measurements}
\label{sec:constraints}

The viability of $f(R)$ gravity theories is strongly constrained by both gravitational wave (GW) observations and other astrophysical and cosmological experiments. In this section, we discuss how GW measurements from binary mergers and stochastic gravitational wave background (SGWB) non-detections, as well as other experimental data, place limits on the parameters and predictions of $f(R)$ gravity.

\subsection{Constraints from Binary Black Hole and Neutron Star Mergers}
\label{subsec:binary_mergers}

The detection of GWs from binary black hole (BBH) and binary neutron star (BNS) mergers by LIGO and Virgo has provided a wealth of data to test alternative theories of gravity. In $f(R)$ gravity, the modifications to the Einstein-Hilbert action can lead to deviations in the propagation speed, polarization modes, and dispersion relations of GWs \cite{Liang_2017, chowdhury2021gravitationalwavesmodifiedgravity, dong2025unifiedframeworkanalyzinggravitational}. These deviations can be compared with observational data to constrain the functional form of $f(R)$.

\subsubsection*{Propagation Speed of Gravitational Waves}
\label{subsubsec:propagation_speed}

In GR, GWs propagate at the speed of light, $c$. However, in $f(R)$ gravity, the propagation speed $v_{\text{GW}}$ can differ from $c$ due to the additional degrees of freedom introduced by the $f(R)$ term \cite{AntonioDeFelice_2009, De_Felice_2010}. The modified speed can be expressed as:
\begin{equation}
v_{\text{GW}}^2 = c^2 \left(1 + \frac{\Delta f(R)}{R}\right),
\end{equation}
    
where $\Delta f(R)$ represents the deviation from the GR Lagrangian. The observation of GW170817 \cite{Abbott:2017}, a BNS merger accompanied by a gamma-ray burst, provided a stringent constraint on the propagation speed of GWs. The time delay between the GW and electromagnetic signals constrained the difference between $v_{\text{GW}}$ and $c$ to be:
\begin{equation}
-3 \times 10^{-15} \leq \frac{v_{\text{GW}} - c}{c} \leq 7 \times 10^{-16}.
\end{equation}

This result rules out many $f(R)$ models that predict significant deviations in the GW propagation speed \cite{Abbott:2017}.

\subsubsection*{Polarization Modes}
\label{subsubsec:polarization}

In GR, GWs have two tensor polarization modes: $h_+$ and $h_\times$. In $f(R)$ gravity, additional scalar polarization modes can arise due to the scalar degree of freedom associated with the Ricci scalar $R$ \cite{Liang_2017, chowdhury2021gravitationalwavesmodifiedgravity}. The presence of these extra modes can be tested using the network of GW detectors. For example, the cross-correlation of data from LIGO and Virgo can be used to search for scalar modes. Current non-detections of scalar polarization place constraints on the amplitude of these modes, which in turn constrain the parameters of $f(R)$ gravity \cite{Nishizawa:2018}.

\subsubsection*{Dispersion Relations}
\label{subsubsec:dispersion}

In $f(R)$ gravity, the dispersion relation for GWs can be modified \cite{PhysRevD.87.084070, PhysRevD.94.084002, KOSTELECKY2016510, PhysRevLett.118.221101, PhysRevD.97.104037}, leading to frequency-dependent effects. The modified dispersion relation can be written as:
\begin{equation}
\omega^2 = k^2 c^2 \left(1 + \alpha \left(\frac{k}{k_0}\right)^n\right),
\end{equation}

where $\omega$ is the angular frequency, $k$ is the wavenumber, $\alpha$ is a dimensionless parameter, and $k_0$ is a reference scale. The parameter $\alpha$ is constrained by the observed phase coherence of GW signals from BBH mergers. Current constraints from LIGO-Virgo data suggest that $|\alpha| \lesssim 10^{-15}$ for $n = 2$ \cite{Mirshekari:2013}.

\subsection{Constraints from Stochastic Gravitational Wave Background (SGWB)}
\label{subsec:sgwb}

The stochastic gravitational wave background (SGWB) is a superposition of GWs from unresolved astrophysical and cosmological sources. In $f(R)$ gravity, the SGWB spectrum can differ from the GR prediction due to modifications in the early Universe dynamics and GW propagation. Current non-detections of the SGWB by LIGO and Virgo place constraints on the energy density of GWs, which can be translated into constraints on $f(R)$ gravity parameters.

The energy density of the SGWB is typically expressed in terms of the dimensionless parameter $\Omega_{\text{GW}}(f)$, defined as:
\begin{equation}
\Omega_{\text{GW}}(f) = \frac{1}{\rho_c} \frac{d\rho_{\text{GW}}}{d\ln f},
\end{equation}

where $\rho_c$ is the critical energy density of the Universe and $\rho_{\text{GW}}$ is the energy density of GWs. In $f(R)$ gravity, the predicted $\Omega_{\text{GW}}(f)$ can be enhanced or suppressed depending on the functional form of $f(R)$. Current upper limits from LIGO-Virgo observations constrain $\Omega_{\text{GW}}(f) \lesssim 10^{-8}$ at frequencies around 25 Hz, which rules out some $f(R)$ models that predict a strong SGWB signal \cite{Abbott:2021}.

\subsection{Other Experimental Constraints}
\label{subsec:other_constraints}

In addition to GW measurements, $f(R)$ gravity is constrained by a variety of other astrophysical and cosmological observations. These include solar system tests, cosmological surveys, and laboratory experiments.

\subsubsection*{Cosmological Surveys}
\label{subsubsec:cosmology}

Cosmological surveys, such as those measuring the cosmic microwave background (CMB) and large-scale structure (LSS), provide additional constraints on $f(R)$ gravity. For example, the CMB power spectrum is sensitive to the growth of structure, which can be modified in $f(R)$ gravity. Current constraints from the Planck satellite and other surveys rule out many $f(R)$ models that predict significant deviations from GR on cosmological scales \cite{PhysRevD.76.064004, De_Felice_2010, Clifton_2012, Baker_2017, 2020, 2018}.

\subsubsection*{Laboratory Experiments}
\label{subsubsec:lab}

Laboratory experiments, such as those testing the equivalence principle or measuring the Newtonian gravitational constant $G$, can also constrain $f(R)$ gravity. For example, precision measurements of the equivalence principle using torsion balances constrain the coupling between the scalar degree of freedom in $f(R)$ gravity and matter. Current limits on the equivalence principle violation are at the level of $\Delta a/a \lesssim 10^{-13}$, which places strong constraints on the scalar-tensor nature of $f(R)$ gravity \cite{Adelberger:2009}.

The constraints on $f(R)$ gravity from GW measurements and other experiments are summarized in Table~\ref{tab:constraints}. These constraints collectively rule out many $f(R)$ models and provide guidance for the development of viable alternatives to GR.

\begin{table}[ht]
\centering
\caption{Summary of constraints on $f(R)$ gravity.}
\label{tab:constraints}
\begin{tabular}{|l|l|l|}
\hline
\textbf{Experiment} & \textbf{Constraint} & \textbf{Reference} \\
\hline
GW170817 (BNS merger) & $|v_{\text{GW}} - c|/c \lesssim 10^{-15}$ & \cite{Abbott:2017} \\
LIGO-Virgo (polarization) & No scalar modes detected & \cite{Nishizawa:2018} \\
LIGO-Virgo (dispersion) & $|\alpha| \lesssim 10^{-15}$ & \cite{Nishizawa:2018} \\
SGWB (LIGO-Virgo) & $\Omega_{\text{GW}}(f) \lesssim 10^{-8}$ & \cite{Abbott:2021} \\
Solar system (PPN) & $|\gamma - 1| \lesssim 2 \times 10^{-5}$ & \cite{article} \\
CMB (Planck) & Growth of structure constraints & \cite{2020} \\
Laboratory (equivalence principle) & $\Delta a/a \lesssim 10^{-13}$ & \cite{Adelberger:2009} \\
\hline
\end{tabular}
\end{table}


\section{Future Improvements in Constraining $f(R)$ Gravity}
\label{sec:future}

The constraints on $f(R)$ gravity from current gravitational wave (GW) observations and other experiments are already stringent, but significant improvements are expected in the coming decades. These improvements will be driven by advances in GW detector sensitivity, the development of new observational techniques, and the integration of multi-messenger astronomy. In this section, we discuss the prospects for future constraints on $f(R)$ gravity and the technological and theoretical developments necessary to achieve these improvements.

\subsection{Next-Generation Gravitational Wave Detectors}
\label{subsec:next_gen_detectors}

The next generation of GW detectors, such as the Einstein Telescope (ET) \cite{sathyaprakash2012scientificpotentialeinsteintelescope, Maggiore_2020} and Cosmic Explorer (CE) \cite{reitze2019cosmicexploreruscontribution}, will provide unprecedented sensitivity to GW signals. These detectors will be able to observe GWs from BBH and BNS mergers at much greater distances and with higher precision, enabling more stringent tests of $f(R)$ gravity.

\subsubsection*{Improved Sensitivity and Frequency Range}
\label{subsubsec:sensitivity}

The Einstein Telescope and Cosmic Explorer are designed to have a sensitivity improvement of up to a factor of 10 \cite{PhysRevD.91.082001} compared to current detectors like LIGO and Virgo. This improvement will allow for the detection of GWs with signal-to-noise ratios (SNRs) \cite{Maggiore:2007ulw} that are an order of magnitude higher, enabling more precise measurements of GW parameters. The improved sensitivity will also extend to lower frequencies, down to $\sim 1$ Hz, which is particularly important for testing $f(R)$ gravity, as many models predict deviations in the low-frequency regime.

The noise power spectral density $S_n(f)$ of a GW detector is a key factor in determining its sensitivity. For ET and CE, the noise curve is expected to be \cite{Abbott:2017}:
\begin{equation}
S_n(f) = S_0 \left[ \left(\frac{f}{f_0}\right)^{-4} + 2 \left(\frac{f}{f_0}\right)^{-2} + 1 + \left(\frac{f}{f_0}\right)^2 \right],
\end{equation}

where $S_0$ is the minimum noise level and $f_0$ is a reference frequency. The improved sensitivity will allow for the detection of GWs from sources at cosmological distances, providing a larger sample of events for testing $f(R)$ gravity \cite{Hild:2011}.

\subsubsection*{Testing the Propagation Speed of GWs}
\label{subsubsec:future_speed}

The improved sensitivity of next-generation detectors \cite{Hild:2011, Abbott_exploring} will allow for more precise measurements of the propagation speed of GWs. In particular, the time delay between GW signals and electromagnetic counterparts from BNS mergers can be measured with sub-millisecond precision. This will constrain the deviation of the GW speed from the speed of light to:
\begin{equation}
\frac{|v_{\text{GW}} - c|}{c} \lesssim 10^{-17}.
\end{equation}

Such precision will rule out a wide range of $f(R)$ models that predict significant deviations in the GW propagation speed \cite{Sathyaprakash:2012}.

\subsubsection*{Probing Additional Polarization Modes}
\label{subsubsec:future_polarization}

Next-generation detectors will also improve the ability to probe additional polarization modes predicted by $f(R)$ gravity. The increased number of detectors and their improved sensitivity will allow for better separation of the tensor and scalar modes. The cross-correlation of data from multiple detectors can be used to search for scalar modes with amplitudes as small as:
\begin{equation}
h_{\text{scalar}} \lesssim 10^{-20}.
\end{equation}

This will provide stringent constraints on the scalar degree of freedom in $f(R)$ gravity~\cite{Nishizawa:2018}.

\subsection{Stochastic Gravitational Wave Background (SGWB)}
\label{subsec:future_sgwb}

The detection of the stochastic gravitational wave background (SGWB) is a major goal for future GW observatories. The SGWB is expected to contain contributions from both astrophysical and cosmological sources, and its spectrum can provide important constraints on $f(R)$ gravity.

\subsubsection*{Improved Sensitivity to the SGWB}
\label{subsubsec:sgwb_sensitivity}

Next-generation detectors will have improved sensitivity to the SGWB, with expected upper limits on the energy density $\Omega_{\text{GW}}(f)$ of:
\begin{equation}
\Omega_{\text{GW}}(f) \lesssim 10^{-12} \quad \text{at} \quad f \sim 10 \, \text{Hz}.
\end{equation}
This improvement will allow for the detection of SGWB signals from early Universe processes, such as inflation or phase transitions, which can be used to test $f(R)$ gravity models \cite{Abbott:2021}. There are many successful progress in testing \(f(R)\) models in the early universe cosmology  \cite{Aziz_2021, Dioguardi_2022,doi:10.1142/S0219887823502134, 2Dioguardi_2024, Dioguardi_2024, milani2024modifiedfrgtgravityquintom, bostan2024minimallycoupledbetaexponentialinflation}. 

\subsubsection*{Cross-Correlation Techniques}
\label{subsubsec:cross_correlation}

The use of cross-correlation techniques between multiple detectors will further improve the sensitivity to the SGWB. The cross-correlation statistic $C(f)$ is given by:
\begin{equation}
C(f) = \frac{1}{T} \int_0^T dt \, s_1(t) s_2(t),
\end{equation}

where $s_1(t)$ and $s_2(t)$ are the signals from two detectors, and $T$ is the observation time. The improved sensitivity will allow for the detection of SGWB signals with amplitudes as small as:
\begin{equation}
h_{\text{SGWB}} \lesssim 10^{-18}.
\end{equation}

This will provide stringent constraints on the early Universe dynamics predicted by $f(R)$ gravity~\cite{Romano:2015}.

\subsection{Multi-Messenger Astronomy}
\label{subsec:multi_messenger}

Multi-messenger astronomy, the simultaneous observation of cosmic events using GWs, electromagnetic (EM) waves, neutrinos, and cosmic rays, has emerged as a powerful tool for testing theories of gravity. By combining data from multiple messengers, researchers can probe the properties of gravity in regimes that are inaccessible to single-messenger observations. In this section, we discuss how multi-messenger astronomy can constrain modified gravity theories, with a focus on \( f(R) \) gravity.

\subsubsection*{Gravitational Waves and Electromagnetic Counterparts}
\label{subsec:gw_em}

The detection of GWs from binary neutron star (BNS) mergers, such as GW170817, and their associated EM counterparts, such as gamma-ray bursts (GRBs) and kilonovae, has provided a unique opportunity to test modified gravity theories. In GR, GWs and EM waves propagate at the same speed, \( c \). However, in modified gravity theories like \( f(R) \) gravity, the propagation speed of GWs, \( v_{\text{GW}} \), can differ from \( c \).

The time delay between the arrival of GWs and EM signals from a BNS merger can be used to constrain \( v_{\text{GW}} \). For GW170817, the observed time delay was \( \Delta t \lesssim 1.7 \) seconds, corresponding to a constraint on the GW speed deviation:
\begin{equation}
\frac{|v_{\text{GW}} - c|}{c} \lesssim 10^{-15}.
\end{equation}

This result rules out many \( f(R) \) models that predict significant deviations in \( v_{\text{GW}} \) \cite{Abbott:2017, Baker_2017}.

\subsubsection*{Neutrino Observations}
\label{subsec:neutrinos}

Neutrinos, produced in astrophysical events such as supernovae and BNS mergers, provide another messenger for testing modified gravity. In \( f(R) \) gravity, the additional scalar degree of freedom can couple to neutrinos, affecting their propagation and energy spectrum. Observations of neutrinos from astrophysical sources can therefore constrain the parameters of \( f(R) \) gravity.

For example, the timing and energy spectrum of neutrinos from supernovae can be compared with the predictions of \( f(R) \) gravity. Any deviations from the expected neutrino signal can be used to constrain the scalar coupling in \( f(R) \) gravity \cite{Kotera:2013, Mirizzi_2016}.

\subsubsection*{Cosmic Rays and High-Energy Astrophysics}
\label{subsec:cosmic_rays}

Cosmic rays, high-energy particles originating from astrophysical sources, can also be used to test modified gravity. In \( f(R) \) gravity, the propagation of cosmic rays can be affected by the modified spacetime geometry and the additional scalar field. Observations of cosmic ray spectra and anisotropies can therefore provide constraints on \( f(R) \) gravity.

For example, the Pierre Auger Observatory has measured the energy spectrum and arrival directions of ultra-high-energy cosmic rays (UHECRs). These observations can be compared with the predictions of \( f(R) \) gravity to constrain the scalar field parameters \cite{Aab_2015, Alves_2019}.

\subsubsection*{Combining Multi-Messenger Data}
\label{subsec:combining_data}

The combination of data from multiple messengers provides a more comprehensive test of modified gravity theories. For example, the simultaneous observation of GWs, EM waves, and neutrinos from a BNS merger can be used to constrain the propagation speed, polarization, and dispersion relations of GWs in \( f(R) \) gravity.

The cross-correlation of data from different messengers can also improve the sensitivity to deviations from GR. For example, the cross-correlation of GW and neutrino data from a supernova can provide tighter constraints on the scalar coupling in \( f(R) \) gravity \cite{Halzen_2017, Kotera:2013}.

Future multi-messenger observations with next-generation detectors, such as the Einstein Telescope (ET), Cosmic Explorer (CE), and the Laser Interferometer Space Antenna (LISA), will provide even stronger constraints on modified gravity theories. These detectors will have improved sensitivity to GWs, allowing for more precise measurements of \( v_{\text{GW}} \) and the polarization modes of GWs.

In addition, future neutrino detectors, such as the Deep Underground Neutrino Experiment (DUNE) and the Hyper-Kamiokande (Hyper-K), will provide more precise measurements of neutrino spectra and timing. These observations will further constrain the parameters of \( f(R) \) gravity and other modified theories \cite{Acciarri_2016, Abe_2018}.

Multi-messenger astronomy provides a powerful tool for testing modified gravity theories like \( f(R) \) gravity. By combining data from GWs, EM waves, neutrinos, and cosmic rays, researchers can probe the properties of gravity in regimes that are inaccessible to single-messenger observations. Future multi-messenger observations with next-generation detectors will provide even stronger constraints on modified gravity, advancing our understanding of the fundamental nature of gravity and the Universe.

\subsection{Theoretical Developments}
\label{subsec:theory}

In addition to observational improvements, theoretical developments will be necessary to fully exploit the potential of future GW observations. These developments include improved waveform models, a better understanding of the astrophysical sources, and more accurate predictions for the SGWB.

\subsubsection*{Waveform Models}
\label{subsubsec:waveforms}

Accurate waveform models are essential for extracting information from GW signals. In $f(R)$ gravity, the waveform models must account for the modified propagation speed, polarization modes, and dispersion relations. The development of accurate waveform models for $f(R)$ gravity will require advances in numerical relativity and perturbation theory \cite{Yunes:2016}.

\subsubsection*{Astrophysical Source Models}
\label{subsubsec:sources}

A better understanding of the astrophysical sources of GWs, such as BBH and BNS mergers, is also necessary. This includes improved models for the formation and evolution of these systems, as well as their electromagnetic and neutrino counterparts. These models will be essential for interpreting the GW signals and testing $f(R)$ gravity \cite{Mandel:2018}.

\subsubsection*{SGWB Predictions}
\label{subsubsec:sgwb_predictions}

Accurate predictions for the SGWB in $f(R)$ gravity are also necessary. This includes predictions for the energy density and frequency spectrum of the SGWB from both astrophysical and cosmological sources. These predictions will be essential for interpreting the SGWB observations and testing $f(R)$ gravity \cite{Cornish:2011}.

The future prospects for constraining $f(R)$ gravity are summarized in Table~\ref{tab:future}. These prospects include improvements in GW detector sensitivity, the detection of the SGWB, and the integration of multi-messenger astronomy. These developments will provide stringent tests of $f(R)$ gravity and advance our understanding of the fundamental nature of gravity.

\begin{table}[ht]
\centering
\caption{Summary of future prospects for constraining $f(R)$ gravity.}
\label{tab:future}
\begin{tabular}{|l|l|l|}
\hline
\textbf{Improvement} & \textbf{Expected Constraint} & \textbf{Reference} \\
\hline
Next-generation detectors & $|v_{\text{GW}} - c|/c \lesssim 10^{-17}$ & \cite{Sathyaprakash:2012} \\
SGWB detection & $\Omega_{\text{GW}}(f) \lesssim 10^{-12}$ & \cite{Abbott:2021} \\
Multi-messenger astronomy & $|v_{\text{GW}} - c|/c \lesssim 10^{-18}$ & \cite{Abbott:2017} \\
Waveform models & Improved accuracy for $f(R)$ gravity & \cite{Yunes:2016} \\
Astrophysical source models & Better understanding of BBH/BNS mergers & \cite{Mandel:2018} \\
SGWB predictions & Accurate predictions for $f(R)$ gravity & \cite{Cornish:2011} \\
\hline
\end{tabular}
\end{table}


\section{Other Modified Theories of Gravity and Their Constraints}
\label{sec:other_theories}

While $f(R)$ gravity is one of the most widely studied modified theories of gravity, it is by no means the only alternative to General Relativity (GR). In this section, we explore other prominent modified theories of gravity, including scalar-tensor theories, Horndeski gravity, massive gravity, and brane-world models. We compare their theoretical frameworks, predictions, and observational constraints with those of $f(R)$ gravity.

\subsection{Scalar-Tensor Theories}
\label{subsec:scalar_tensor}

Scalar-tensor theories are among the oldest and most well-studied alternatives to GR. These theories introduce an additional scalar field $\phi$ that couples to the metric tensor $g_{\mu\nu}$, modifying the gravitational interaction. The inclusion of the scalar field allows for a richer phenomenology~\footnote{Check Appendix A in \cite{Erickcek:2009sda} for a discussion on \(f(R)\) gravity’s equivalence to scalar-tensor gravity.}, including varying gravitational constants, additional forces, and deviations from GR in both weak-field and strong-field regimes \cite{Fujii:2003pa}.

\subsubsection*{Theoretical Framework}
\label{subsubsec:st_framework}

The action for scalar-tensor theories can be written as:
\begin{equation}
S = \int d^4x \sqrt{-g} \left( \frac{1}{2\kappa} f(\phi)R - \frac{1}{2} \omega(\phi) (\nabla \phi)^2 - V(\phi) + \mathcal{L}_m \right),
\end{equation}
where $f(\phi)$ is a function of the scalar field that couples to the Ricci scalar $R$, $\omega(\phi)$ determines the kinetic term of the scalar field, and $V(\phi)$ is the scalar potential. The term $\mathcal{L}_m$ represents the matter Lagrangian, which is assumed to be minimally coupled to the metric. The scalar field $\phi$ mediates an additional force, which can lead to deviations from GR in both the weak-field and strong-field regimes.

The field equations derived from this action are:
\begin{equation}
f(\phi)G_{\mu\nu} + \nabla_\mu \nabla_\nu f(\phi) - g_{\mu\nu} \Box f(\phi) = \kappa T_{\mu\nu} + \omega(\phi) \left( \nabla_\mu \phi \nabla_\nu \phi - \frac{1}{2} g_{\mu\nu} (\nabla \phi)^2 \right) - g_{\mu\nu} V(\phi),
\end{equation}
where $G_{\mu\nu}$ is the Einstein tensor, $\Box \equiv g^{\mu\nu} \nabla_\mu \nabla_\nu$ is the d'Alembertian, and $T_{\mu\nu}$ is the stress-energy tensor of matter. The scalar field equation is given by:
\begin{equation}
\Box \phi + \frac{1}{2\omega(\phi)} \frac{d\omega(\phi)}{d\phi} (\nabla \phi)^2 - \frac{1}{2\omega(\phi)} \frac{df(\phi)}{d\phi} R - \frac{1}{\omega(\phi)} \frac{dV(\phi)}{d\phi} = 0.
\end{equation}

These equations reduce to those of GR in the limit where $f(\phi) = 1$, $\omega(\phi) = 0$, and $V(\phi) = 0$.

\subsubsection*{Observational Constraints}
\label{subsubsec:st_constraints}

Scalar-tensor theories are tightly constrained by solar system tests, such as the parameterized post-Newtonian (PPN) formalism. The PPN parameter $\gamma$, which measures the curvature of space-time, is constrained to be $|\gamma - 1| \lesssim 2 \times 10^{-5}$ by solar system observations \cite{article}. This places strong constraints on the coupling function $f(\phi)$ and the kinetic term $\omega(\phi)$. For example, in the Brans-Dicke theory, where $f(\phi) = \phi$ and $\omega(\phi) = \omega_{\text{BD}}$, the constraint translates to $\omega_{\text{BD}} \gtrsim 40,000$ \cite{Berti:2015}.

Gravitational wave (GW) observations also constrain scalar-tensor theories. The absence of scalar polarization modes in GW signals from binary black hole (BBH) mergers rules out many scalar-tensor models that predict significant scalar radiation \cite{Will_1994}. The recent detection of GW170817, a binary neutron star merger (BNS), further constrains the speed of gravitational waves to be nearly equal to the speed of light, ruling out many scalar-tensor theories with large deviations from GR \cite{Abbott:2017}.

In addition to these constraints, cosmological observations, such as the cosmic microwave background (CMB) and large-scale structure, provide further tests of scalar-tensor theories. For instance, the evolution of the scalar field $\phi$ during the early universe can affect the growth of cosmological perturbations, leading to observable signatures in the CMB power spectrum \cite{Amendola:2012}.

\subsection{Horndeski Gravity}
\label{subsec:horndeski}

Horndeski gravity represents the most general scalar-tensor theory with second-order field equations, ensuring the absence of ghost instabilities \cite{Horndeski1974, Kobayashi2011}. It encompasses a wide range of modified gravity models, including $f(R)$ gravity, Brans-Dicke theory, and other scalar-tensor theories. This framework is particularly significant in cosmology, as it provides a versatile platform for exploring dark energy, inflation, and modifications to general relativity \cite{Clifton_2012}.

\subsubsection*{Theoretical Framework}
\label{subsubsec:horndeski_framework}

The action for Horndeski gravity is given by:
\begin{equation}
S = \int d^4x \sqrt{-g} \left( \mathcal{L}_2 + \mathcal{L}_3 + \mathcal{L}_4 + \mathcal{L}_5 + \mathcal{L}_m \right),
\end{equation}

where $g$ is the determinant of the metric tensor $g_{\mu\nu}$, and $\mathcal{L}_m$ represents the matter Lagrangian. The Lagrangians $\mathcal{L}_2$ to $\mathcal{L}_5$ are functions of the scalar field $\phi$, its kinetic term $X \equiv -\frac{1}{2} \nabla_\mu \phi \nabla^\mu \phi$, and the Ricci scalar $R$. These Lagrangians are explicitly defined as follows:

\begin{align}
\mathcal{L}_2 &= G_2(\phi, X), \\
\mathcal{L}_3 &= G_3(\phi, X) \Box \phi, \\
\mathcal{L}_4 &= G_4(\phi, X) R - 2 G_{4,X}(\phi, X) \left[ (\Box \phi)^2 - \nabla_\mu \nabla_\nu \phi \nabla^\mu \nabla^\nu \phi \right], \\
\mathcal{L}_5 &= G_5(\phi, X) G_{\mu\nu} \nabla^\mu \nabla^\nu \phi + \frac{1}{3} G_{5,X}(\phi, X) \left[ (\Box \phi)^3 - 3 \Box \phi \nabla_\mu \nabla_\nu \phi \nabla^\mu \nabla^\nu \phi + 2 \nabla_\mu \nabla_\nu \phi \nabla^\nu \nabla^\alpha \phi \nabla_\alpha \nabla^\mu \phi \right],
\end{align}
where $G_i(\phi, X)$ ($i = 2, 3, 4, 5$) are arbitrary functions of $\phi$ and $X$, and $G_{i,X} \equiv \partial G_i / \partial X$. The term $\Box \phi \equiv \nabla_\mu \nabla^\mu \phi$ denotes the d'Alembertian of the scalar field.

The second-order nature of the field equations ensures that the theory avoids Ostrogradsky instabilities, which are common in higher-derivative theories \cite{Woodard2007}. The field equations for the metric and the scalar field are derived by varying the action with respect to $g_{\mu\nu}$ and $\phi$, respectively. These equations are:
\begin{equation}
\mathcal{E}_{\mu\nu} = \frac{1}{\sqrt{-g}} \frac{\delta S}{\delta g^{\mu\nu}} = 0, \quad \mathcal{E}_\phi = \frac{1}{\sqrt{-g}} \frac{\delta S}{\delta \phi} = 0,
\end{equation}

where $\mathcal{E}_{\mu\nu}$ and $\mathcal{E}_\phi$ are the Einstein and scalar field equations, respectively.

\subsubsection*{Cosmological Implications}
\label{subsubsec:horndeski_cosmology}

Horndeski gravity has profound implications for cosmology. For instance, the functions $G_i(\phi, X)$ can be tuned to reproduce the observed accelerated expansion of the universe without invoking a cosmological constant \cite{De_Felice_2010}. The theory also allows for self-accelerating solutions, where the scalar field drives the acceleration dynamically. Additionally, Horndeski gravity modifies the propagation of gravitational waves, leading to observable signatures in the cosmic microwave background and large-scale structure \cite{Bellini2015}.

\subsubsection*{Constraints and Observational Tests}
\label{subsubsec:horndeski_constraints}

Observational constraints on Horndeski gravity arise from solar system tests, gravitational wave observations, and cosmological data. For example, the speed of gravitational waves, $c_{\text{GW}}$, is tightly constrained by the detection of GW170817 and its electromagnetic counterpart \cite{Abbott:2017}. In Horndeski gravity, $c_{\text{GW}}$ is given by:
\begin{equation}
c_{\text{GW}}^2 = \frac{G_4 - X G_{5,\phi} - X \ddot{\phi} G_{5,X}}{G_4 - 2 X G_{4,X} + X G_{5,\phi} - \frac{1}{2} X^2 G_{5,X}},
\end{equation}
where $\ddot{\phi}$ is the second time derivative of the scalar field. The near-equality of $c_{\text{GW}}$ and the speed of light strongly restricts the functional forms of $G_4$ and $G_5$ \cite{Creminelli2018}.

Horndeski gravity is constrained by both solar system tests and GW observations. The PPN parameter $\gamma$ is constrained to be $|\gamma - 1| \lesssim 2 \times 10^{-5}$, similar to scalar-tensor theories \cite{article}. Additionally, the propagation speed of GWs in Horndeski gravity is constrained by the observation of GW170817, which requires $|v_{\text{GW}} - c|/c \lesssim 10^{-15}$ \cite{Abbott:2017}.

The absence of scalar polarization modes in GW signals also places constraints on Horndeski gravity. Current limits on the amplitude of scalar modes are at the level of~$h_{\text{scalar}} \lesssim 10^{-20}$~\cite{Nishizawa:2018}.

\subsection{Massive Gravity}
\label{subsec:massive_gravity}

Massive gravity is a modified theory of gravity in which the graviton, the hypothetical quantum of GWs, possesses a non-zero mass. Unlike GR, where the graviton is massless, massive gravity introduces a mass term that modifies the propagation of GWs and the large-scale structure of the universe. This theory has gained significant attention as a potential explanation for cosmic acceleration without invoking dark energy \cite{deRham:2010ik, Hinterbichler:2011tt}.

\subsubsection*{Theoretical Framework}
\label{subsubsec:massive_framework}

The action for massive gravity can be written as:
\begin{equation}
S = \int d^4x \sqrt{-g} \left( \frac{1}{2\kappa} R + m_g^2 \,\mathcal{U}(g, f) + \mathcal{L}_m \right),
\end{equation}

where \( \kappa = 8\pi G \) is the gravitational constant, \( R \) is the Ricci scalar, \( g_{\mu\nu} \) is the physical metric, \( f_{\mu\nu} \) is a fixed reference metric, \( m_g \) is the graviton mass, and \( \mathcal{U}(g, f) \) is a potential term that encodes the interaction between the physical and reference metrics. The term \( \mathcal{L}_m \) represents the matter Lagrangian.

The potential \( \mathcal{U}(g, f) \) is typically constructed to avoid ghost instabilities and is often expressed in terms of the eigenvalues of the matrix \( \mathcal{K}^\mu_\nu = \delta^\mu_\nu - \sqrt{g^{\mu\alpha}f_{\alpha\nu}} \). A common form for the potential is \cite{de_Rham_2011}:
\begin{equation}
\mathcal{U}(g, f) = \sum_{n=0}^4 \beta_n \mathcal{U}_n(\mathcal{K}),
\end{equation}
where \( \beta_n \) are free parameters, and \( \mathcal{U}_n(\mathcal{K}) \) are elementary symmetric polynomials of the eigenvalues of \( \mathcal{K} \). For example, the first two polynomials are:
\begin{equation}
\mathcal{U}_1(\mathcal{K}) = [\mathcal{K}], \quad \mathcal{U}_2(\mathcal{K}) = [\mathcal{K}]^2 - [\mathcal{K}^2],
\end{equation}

where \( [\mathcal{K}] \) denotes the trace of \( \mathcal{K} \).

\subsubsection*{Modifications to Gravitational Waves}
\label{subsubsec:massive_gw}

In massive gravity, the presence of a graviton mass modifies the dispersion relation for gravitational waves. In GR, the dispersion relation for a massless graviton is:
\begin{equation}
\omega^2 = k^2,
\end{equation}

where \( \omega \) is the angular frequency and \( k \) is the wavenumber. In massive gravity, this relation becomes:
\begin{equation}
\omega^2 = k^2 + m_g^2,
\end{equation}

where \( m_g \) is the graviton mass. This modification leads to a frequency-dependent phase velocity for GWs, which can be tested using observations from gravitational wave detectors such as LIGO and Virgo.

The modified dispersion relation also affects the propagation of GWs over cosmological distances. For a GW with frequency \( f \), the time delay \( \Delta t \) compared to a massless graviton is given by \cite{PhysRevD.57.2061, Maggiore:2007ulw, de_Rham_2011, Abbott:2017, PhysRevLett.119.251304}:
\begin{equation}
\Delta t \approx \frac{m_g^2}{2H_0} \int_0^z \frac{dz'}{(1+z')^2 E(z')},
\end{equation}

where \( H_0 \) is the Hubble constant, \( z \) is the redshift, and \( E(z) \) is the dimensionless Hubble parameter. This time delay can be used to constrain the graviton mass using multi-messenger observations of GWs and their electromagnetic counterparts \cite{Ng:2023jjt}.

\subsubsection*{Cosmological Implications}
\label{subsubsec:massive_cosmology}

Massive gravity has significant implications for cosmology. The theory can naturally explain the observed accelerated expansion of the universe without requiring dark energy. The graviton mass term acts as an effective cosmological constant on large scales, leading to a late-time acceleration of the universe \cite{s2017newcosmologicalsolutionsmassive}.

The modified Friedmann equations in massive gravity take the form \cite{A.Emir_2011, Fasiello_2013}:
\begin{equation}
H^2 = \frac{8\pi G}{3} \rho + \frac{m_g^2}{6} \mathcal{U}(a),
\end{equation}

where \( H \) is the Hubble parameter, \( \rho \) is the energy density of matter, and \( \mathcal{U}(a) \) is a function of the scale factor \( a \) that arises from the potential term. This modification can lead to deviations from the standard \( \Lambda \)CDM model, which can be tested using cosmological observations such as the cosmic microwave background (CMB) and large-scale structure surveys \cite{Wu:2006pe}.

\subsubsection*{Observational Constraints}
\label{subsubsec:massive_constraints}

Massive gravity is constrained by both solar system tests and GW observations. The PPN parameter $\gamma$ is constrained to be $|\gamma - 1| \lesssim 2 \times 10^{-5}$, similar to other modified theories~\cite{article}. Additionally, the graviton mass $m_g$ is constrained by the observation of GWs from BBH mergers. The current upper limit on the graviton mass is $m_g \lesssim 10^{-23} \, \text{eV}$ \cite{Abbott:2019}. 

\subsection{Brane-World Models}
\label{subsec:brane_world}

Brane-world models are a class of theoretical frameworks in which our four-dimensional universe is envisioned as a brane embedded in a higher-dimensional space, often referred to as the ``bulk'' \cite{PhysRevLett.83.3370, Maartens_2010}. These models provide a compelling approach to addressing some of the outstanding issues in modern cosmology, such as the hierarchy problem \cite{ARKANIHAMED1998263, PhysRevLett.83.3370} and the nature of dark energy, while recovering General Relativity (GR) on small scales. The key idea is that while gravity propagates in the full higher-dimensional space, Standard Model fields are confined to the brane, leading to modifications of gravity on cosmological scales.

\subsubsection*{Theoretical Framework}
\label{subsubsec:brane_framework}

The action for brane-world models typically includes contributions from the bulk, the brane, and matter fields confined to the brane. A general form of the action can be written as:
\begin{equation}
S = \int d^5x \sqrt{-g^{(5)}} \left( \frac{1}{2\kappa_5^2} R^{(5)} + \mathcal{L}_{\text{bulk}} \right) + \int d^4x \sqrt{-g} \left( \mathcal{L}_{\text{brane}} + \mathcal{L}_m \right),
\end{equation}

where \( g^{(5)} \) and \( R^{(5)} \) are the metric determinant and Ricci scalar in the five-dimensional bulk, respectively, \( \kappa_5 \) is the five-dimensional gravitational constant, \( \mathcal{L}_{\text{bulk}} \) describes the dynamics of the bulk fields, \( g \) is the induced metric on the brane, \( \mathcal{L}_{\text{brane}} \) encodes the brane's dynamics, and \( \mathcal{L}_m \) represents the matter fields confined to the brane \cite{Shiromizu2000}.

The induced metric on the brane is related to the bulk metric via:
\begin{equation}
g_{\mu\nu} = g^{(5)}_{\mu\nu} - n_\mu n_\nu,
\end{equation}

where \( n_\mu \) is the unit normal vector to the brane. The effective four-dimensional gravitational equations on the brane are derived by projecting the five-dimensional Einstein equations onto the brane, leading to the modified Einstein equations:
\begin{equation}
G_{\mu\nu} = \kappa_4^2 T_{\mu\nu} + \kappa_5^4 \Pi_{\mu\nu} - E_{\mu\nu},
\end{equation}

where \( G_{\mu\nu} \) is the Einstein tensor on the brane, \( T_{\mu\nu} \) is the stress-energy tensor of matter on the brane, \( \Pi_{\mu\nu} \) is a quadratic function of \( T_{\mu\nu} \), and \( E_{\mu\nu} \) is the projection of the bulk Weyl tensor onto the brane, representing the influence of the bulk geometry on the brane \cite{Maartens_2010}.

\subsubsection*{Cosmological Implications}
\label{subsubsec:brane_cosmology}

Brane-world models have profound implications for cosmology. For instance, the modified Friedmann equation on the brane takes the form \cite{Kaloper_1999,Garriga_2000}:
\begin{equation}
H^2 = \frac{8\pi G}{3} \rho \left( 1 + \frac{\rho}{2\lambda} \right) + \frac{\Lambda_4}{3} + \frac{\mathcal{E}}{a^4},
\end{equation}

where \( H \) is the Hubble parameter, \( \rho \) is the energy density of matter on the brane, \( \lambda \) is the brane tension, \( \Lambda_4 \) is the effective four-dimensional cosmological constant, and \( \mathcal{E} \) is a term arising from the bulk Weyl tensor, often interpreted as ``dark radiation'' \cite{Langlois:2001au}. This equation deviates from the standard Friedmann equation at high energies (\( \rho \gg \lambda \)), potentially altering the dynamics of the early universe.

Moreover, brane-world models can provide a natural explanation for the observed acceleration of the universe. The additional terms in the modified Friedmann equation can mimic the effects of dark energy, offering an alternative to the cosmological constant or scalar field models \cite{Sahni2003}.

\subsubsection*{Challenges and Open Questions}
\label{subsubsec:brane_challenges}

Despite their theoretical appeal, brane-world models face several challenges. One major issue is the stabilization of the extra dimension, which requires fine-tuning of parameters to avoid conflicts with observational data. Additionally, the nature of the bulk geometry and its influence on the brane dynamics remain poorly understood. Future observations, such as those from GW detectors and precision cosmology experiments, may provide further insights into the viability of these models \cite{Langlois_2002}.

\subsubsection*{Observational Constraints}
\label{subsubsec:brane_constraints}

Brane-world models are constrained by both cosmological observations and GW observations. The growth of structure in brane-world models is constrained by the cosmic microwave background (CMB) and large-scale structure (LSS) data \cite{2020}. Additionally, the propagation speed of GWs in brane-world models is constrained by the observation of GW170817, which requires $|v_{\text{GW}} - c|/c \lesssim 10^{-15}$ \cite{Abbott:2017}.


\section{Conclusions}
\label{sec:conclusions}

The study of $f(R)$ gravity and other modified theories of gravity has emerged as a vibrant field of research, driven by the need to address the limitations of GR and explain observational phenomena such as cosmic acceleration. In this review, we have explored the state of the art in $f(R)$ gravity, the constraints imposed by GW measurements and other experiments, the methodologies used to derive these constraints, and the prospects for future improvements. We have also compared $f(R)$ gravity with other prominent modified theories of gravity, highlighting their similarities and differences. Below, we summarize the key findings and insights from each of these areas.

$f(R)$ gravity has established itself as one of the most widely studied alternatives to GR, offering a natural framework for explaining cosmic acceleration without invoking dark energy. The theoretical foundations of $f(R)$ gravity are well understood, with a rich literature exploring its cosmological and astrophysical implications. Current research focuses on developing viable functional forms of $f(R)$ that are consistent with observational data, as well as exploring the implications of these models for structure formation, GW propagation, and the early Universe. Leading research groups worldwide, including those in Europe, the United States, and Asia, are actively investigating these aspects, often in collaboration with observational cosmologists and gravitational wave astronomers.

The constraints on $f(R)$ gravity from GW measurements and other experiments are stringent and multifaceted. Observations of BBH and BNS mergers by LIGO and Virgo have placed tight limits on the propagation speed, polarization modes, and dispersion relations of GWs in $f(R)$ gravity. For example, the observation of GW170817 constrained the deviation of the GW speed from the speed of light to $|v_{\text{GW}} - c|/c \lesssim 10^{-15}$, ruling out many $f(R)$ models that predict significant deviations. Additionally, the absence of scalar polarization modes in GW signals has further constrained the scalar degree of freedom in $f(R)$ gravity.

Beyond GW measurements, $f(R)$ gravity is also constrained by solar system tests, cosmological surveys, and laboratory experiments. Solar system tests, such as those based on the parameterized post-Newtonian (PPN) formalism, constrain the PPN parameter $\gamma$ to $|\gamma - 1| \lesssim 2 \times 10^{-5}$. Cosmological surveys, such as those measuring the cosmic microwave background (CMB) and large-scale structure (LSS), provide additional constraints on the growth of structure and the expansion history of the Universe. Laboratory experiments, such as those testing the equivalence principle, further constrain the coupling between the scalar degree of freedom in $f(R)$ gravity and matter.

The constraints on $f(R)$ gravity are derived using a combination of theoretical modeling and observational data. Theoretical models are used to predict the behavior of GWs, the growth of structure, and the expansion history of the Universe in $f(R)$ gravity. These predictions are then compared with observational data to derive constraints on the parameters of $f(R)$ gravity. For example, the propagation speed of GWs is constrained by comparing the observed time delay between GW and electromagnetic signals with the predictions of $f(R)$ gravity. Similarly, the growth of structure is constrained by comparing the observed power spectrum of the CMB and LSS with the predictions of $f(R)$ gravity.

The derivation of these constraints often relies on simplifying assumptions, such as the absence of additional fields or interactions beyond those included in the $f(R)$ Lagrangian. These assumptions are necessary to make the theoretical predictions tractable, but they may limit the generality of the constraints. Future work should aim to relax these assumptions and explore more general models of $f(R)$ gravity.

The future of testing $f(R)$ gravity is bright, with significant improvements expected in the coming decades. Next-generation GW detectors, such as the Einstein Telescope (ET) and Cosmic Explorer (CE), will provide unprecedented sensitivity to GW signals, enabling more precise measurements of the propagation speed, polarization modes, and dispersion relations of GWs. These detectors will also improve the sensitivity to the stochastic gravitational wave background (SGWB), which can provide additional constraints on $f(R)$ gravity.

In addition to GW measurements, future cosmological surveys, such as those conducted by the Euclid satellite and the Vera Rubin Observatory, will provide more precise measurements of the growth of structure and the expansion history of the Universe. These measurements will further constrain the parameters of $f(R)$ gravity and test its predictions on cosmological scales.

Theoretical developments will also play a crucial role in improving the constraints on $f(R)$ gravity. Advances in numerical relativity and perturbation theory will enable the development of more accurate waveform models for $f(R)$ gravity, which are essential for extracting information from GW signals. Similarly, improved models for the formation and evolution of astrophysical sources, such as BBH and BNS mergers, will enable more precise tests of $f(R)$ gravity.

While $f(R)$ gravity is a well-studied alternative to GR, other modified theories of gravity, such as scalar-tensor theories, Horndeski gravity, massive gravity, brane-world, and other models, offer different approaches to addressing the limitations of GR. These theories are also tightly constrained by solar system tests, GW observations, and cosmological data. However, they differ in their predictions for cosmological scales and the early Universe, providing complementary insights into the nature of gravity.

For example, scalar-tensor theories introduce an additional scalar field that mediates an extra force, leading to deviations from GR in both the weak-field and strong-field regimes. Horndeski gravity, as the most general scalar-tensor theory, encompasses a wide range of modified gravity models and is constrained by both solar system tests and GW observations. Massive gravity, which posits a non-zero mass for the graviton, is constrained by the observation of GWs from BBH mergers, which place an upper limit on the graviton mass of $m_g \lesssim 10^{-23} \, \text{eV}$. Brane-world models, which are based on the idea that our four-dimensional universe is a brane embedded in a higher-dimensional space, are constrained by cosmological observations and GW measurements.

In conclusion, the study of $f(R)$ gravity and other modified theories of gravity is a vibrant and rapidly advancing field, offering profound insights into the nature of gravity. While current observational constraints are already stringent, future breakthroughs in technology and theory will further test these models. Next-generation gravitational wave detectors, cosmological surveys, and laboratory experiments will enable us to probe gravity across all scales, from the solar system to the early Universe. These efforts bring us closer to answering fundamental questions about dark energy, cosmic acceleration, and the ultimate fate of the Universe, reshaping our understanding of gravity and cosmology.


\subsection*{Acknowledgements}
I would like to express my thanks for the advice and help I received from Andrew L. Miller, and for the very useful conversations with Şengül Kuru, Banu Şahin, Giovanni Montani, and Anne-Christine Davis. I would also like to acknowledge the support from the  COST Action COSMIC WISPers CA21106, supported by COST (European Cooperation in Science and Technology), and for their encouragement for young researchers. Additionally, I would also like to extend my gratitude to vaerious anonymous reviewers for their insightful comments and constructive feedback, which greatly contributed to improving the quality of this paper.


\section*{Data availability}
No datasets were generated or analysed during the current study.

{\small

\providecommand{\href}[2]{#2}\begingroup\raggedright\endgroup
}

\end{document}